      [1999/12/01 v1.4c Il Nuovo Cimento]
\documentclass{cimento}
\usepackage{graphicx}  

\newcommand{\lesssim}{\lower.5ex\hbox{$\; \buildrel < \over\sim \;$}}
\newcommand{\gtrsim}{\lower.5ex\hbox{$\; \buildrel > \over\sim \;$}}
\newcommand{\e}{\epsilon}

\newcommand{\ep}{\epsilon^\prime}
\newcommand{\gp}{\gamma^\prime}
\newcommand{\g}{\gamma}

\newcommand{\tp}{t^\prime}
\newcommand{\G}{\Gamma}

\title{External Shocks, UHECRs, and 
the Early Afterglow of GRBs}
\author{Charles D.~Dermer\from{ins:x}}
\instlist{\inst{ins:x} E. O. Hulburt Center for Space Research, Code 7653,
Naval Research Laboratory, Washington, D.C. 20375-5352; 
\small dermer@gamma.nrl.navy.mil }
\PACSes{
\PACSit{95.85.Ry}{98.70.Rz}}
\begin{document}

\maketitle

\begin{abstract}

Highly variable $\gamma$-ray pulses and X-ray flares in GRB light
curves can result from external shocks
rather than central engine activity under the assumption 
that the GRB blast-wave
shell does not spread.  
Acceleration of cosmic rays to $\gtrsim
10^{20}$ eV energies can take place in the external 
shocks of GRBs. 
Escape of hadronic energy in the form of UHECRs 
leads to a rapidly decelerating
GRB blast wave, which may account for the  rapid X-ray declines
observed in  Swift GRBs.
\end{abstract}

\section{Introduction}

GRB light curves measured with Swift consist 
 of a BAT light curve in the 15 -- 150 keV range
followed, after slewing within $\approx 100 $ s, by a detailed 0.3 -- 10
keV XRT X-ray light curve \cite{geh04}.
This information supplements our knowledge of the highly variable
hard X-ray and $\gamma$-ray light curves measured from many GRBs with
BATSE and other GRB detectors.  About one-half of Swift GRBs show
X-ray flares or short timescale structure, sometimes hours or later
after the onset of the GRB.
Approximately $30$\% of the Swift
GRBs display rapid X-ray
declines, and an additional $\approx 30$\% display features unlike
simple blast wave model predictions \cite{obr06}.

We make three points in this paper:
\begin{enumerate}
\item Highly variable light curves can be produced by an external 
shock under the assumption that the GRB blast wave does not
spread, or spreads much more slowly than assumed from gas-dynamic
or relativistic hydrodynamic models that do not take into account
magnetic effects in GRB blast waves. If this assumption
is valid, then it is wrong to conclude that highly variable $\gamma$-ray emissions,
X-ray flares with $\Delta t/t\ll 1$, or late time X-ray flares require
delayed central engine activity or colliding shells. 
\item External shocks in GRB blast waves can accelerate cosmic ray protons
and ions to $\gtrsim 10^{20}$ eV, making GRBs a logical candidate to 
accelerate the highest energy cosmic rays.
\item Escape of ultra-high energy cosmic rays (UHECRs) takes place
from an external shock formed by an expanding GRB blast wave
on time scales of a few hundred seconds
for the observer. Blast-wave deceleration due to the loss of the 
internal hadronic energy is proposed \cite{der06b} to be the cause
of X-ray declines in GRB light curves observed with Swift.
\end{enumerate}

\section{X-ray flares and $\gamma$-ray pulses from external shocks}

We have performed a detailed analysis of the interaction between
a GRB blast-wave shell and an external stationary cloud \cite{der06a}.
The analysis is performed under the assumption that
the cloud width $\Delta_{cl} \ll x$, where $x$ is the distance of the
cloud from the GRB explosion. The interaction is divided into three
phases: (1) a collision phase with both a forward and reverse shock;
(2) a penetration phase where either the reverse shock has crossed the
shell while the forward shock continues to cross the cloud, or vice
versa; and (3) an expansion phase, where both shocks have crossed the
cloud and shell, and the shocked fluid expands. The shell width is written
as 
\begin{equation} \Delta(x) \cong
 \Delta_0 + \eta\;{x\over \Gamma_0^2}\;, 
\label{Delta(x)}
\end{equation}
and the proper number density of the relativistic shell is given by
\begin{equation}
n(x) = {E_0 \over 4\pi x^2 \Gamma_0^2 m_p c^2 \Delta(x)} \;,
\label{n(x)}
\end{equation}
where $\Gamma_0$ is the coasting Lorentz factor of the GRB blast wave, and $E_0$ is 
the apparent isotropic energy release.

Short timescale flaring requires
(a) a strong forward shock, which from the relativistic shock jump conditions \cite{bm76}
imply a maximum cloud density given by
\begin{equation}
n_{cl}  \lesssim {E_0 \over 16\pi x^2 \Gamma_0^4 m_p c^2 \Delta(x)}\;,
\label{nclmax}
\end{equation}
and (b) significant blast-wave deceleration to provide efficient energy
extraction, which occurs in clouds with thick columns
\cite{dm99}, that is, with densities
\begin{equation}
n_{cl}  \gtrsim {E_0 \over 4\pi x_0^2 \Gamma_0^2 m_p c^2 \Delta_{cl}}\;.
\label{nclmin}
\end{equation}
These two conditions 
translate into the requirement that
\begin{equation}
\Delta_{cl} \gtrsim 4\G_0^2 \Delta(x)\;
\label{deltacl}
\end{equation}
in order to produce short timescale variability. The short timescale
variabilty condition \cite{dm99} for quasi-spherical clouds is
\begin{equation}
 \Delta_{cl} \lesssim {x\over \G_0} \;.
\label{Deltacl}
\end{equation} 
 
Using eq.\ (\ref{Delta(x)}) for the shell width, eqs.\ (\ref{deltacl}) and 
(\ref{Deltacl}) imply the requirement that
\begin{equation}
\eta\lesssim 1/4\Gamma_0
\label{eta}
\end{equation}
in order to produce rapid variability from an external shock.
Hence the production of  $\gamma$-ray pulses and 
X-ray flares from external shocks depends on 
whether the GRB blast-wave width spreads in the coasting
phase according to eq.\ (\ref{Delta(x)}),
with $\eta \approx 1$, as is generally argued.  In the gas-dynamical study of \cite{mlr93},
inhomogeneities in the GRB fireball produce a spread
in particle velocities of order $|v - c|/c \sim \Gamma_0^{-2}$, 
so that $\Delta(x) \sim x/\Gamma_0^2$ when
$x \gtrsim \Gamma_0^2 \Delta_0$. This dependence is also obtained in a 
hydrodynamical analysis \cite{psn93}.

Two points can be made about these relations. First,
the spread in $\Delta$ considered for 
a spherical fireball is averaged over all directions. 
As the fireball expands and becomes transparent, the variation
in fluid motions or gas particle directions over a small 
solid angle $\sim 1/\Gamma_0^2$ of the full sky becomes substantially
less. Second, the particles within a magnetized blast-wave shell will
expand and adiabatically cool so that the fluid will spread
with thermal speed $v_{th}= \beta_{th} c$. The comoving width of the blast
wave is $\Gamma_0 \Delta_0 + \beta_{th} c \Delta \tp \approx 
\Gamma_0 \Delta_0 + \beta_{th} x /\Gamma_0$, so that the 
spreading radius $x_{spr} \cong \Gamma_0^2 \Delta_0/\beta_{th}$.
Adiabatic expansion of nonrelativistic particles 
can produce a very cold shell with
$\beta_0 \lesssim 10^{-3}$, leading to very small shell 
widths. 

The requirement on the thinness of $\Delta(x)$ does not 
apply to the adiabatic self-similar phase, where
the width is necessarily $\sim x/\Gamma_0^2$, as implied
by the relativistic shock hydrodynamic equations \cite{bm76}.
Even in this case, however, $\Delta \ll x/\Gamma_0^2$
if the blast wave is highly radiative \cite{cps98}.
Under the 
assumption of a strong forward shock and small clouds in 
the vicinity of a GRB, highly variable GRB light 
curves are formed with reasonable efficiency ($\gtrsim 10$\%) to transform
blast wave energy into $\gamma$ rays
\cite{dm99,dm04}.

\section{Cosmic ray acceleration in GRB blast waves}

The maximum particle energy for a cosmic ray proton accelerated
by an external shock in a GRB blast wave is derived. 
Consider a GRB blast wave with apparent isotropic energy release 
$E_0 = 10^{54}E_{54}$ ergs, (initial) coasting Lorentz factor $\Gamma_0 = 
300\Gamma_{300}$, and external medium density $n_0 = 100n_2$ cm$^{-3}$.
The comoving blast wave volume for the assumed spherically symmertric
explosion, after reaching distance $x$ from the center of the explosion, is 
\begin{equation}
V^\prime = 4\pi x^2 \Delta^\prime,
\end{equation}
where the shell width $\Delta^\prime = x/12\Gamma$ (the factor $1/12 \Gamma$
is the product of the geometrical factor $1/3$ and the factor $1/4\Gamma$ from
the continuity equations of relativistic hydrodynamics; $\Gamma$ is the 
evolving GRB blast wave Lorentz factor). 

The Hillas condition \cite{hil84} for maximum particle energy $E^\prime_{max}$
is that the particle Larmor radius is less than 
the size scale of the system; $E_{max}$
in the stationary frame (primes refer to the comoving frame) is given by 
\begin{equation}
r_L^\prime = {E_{max}^\prime\over eB^\prime } = {E_{max}\over \Gamma e B^\prime } < \Delta^\prime.
\end{equation}
The largest particle energy is reached at the deceleration radius $x = x_d$ when $\Gamma \cong \Gamma_0$, 
where the deceleration radius
\begin{equation}
x_d \equiv ( {3 E_0\over 4\pi\Gamma_0^2 m_p n_0})^{1/3} \cong 2.6\times 10^{16} ({E_{54}\over \Gamma_{300}^2 n_2})^{1/3}\;\rm{cm}\;.
\end{equation}
Hence $
E_{max} \cong {Z eB^\prime x_d/ 12}\;.
$

The mean magnetic field $B^\prime$ in the GRB blast wave
is  assigned in terms of a magnetic field parameter $\e_B$ 
that gives the magnetic field energy density in terms of the 
energy density of the downstream shocked fluid, so 
\begin{equation}
B^\prime = (32\pi n_0 \e_B m_pc^2)^{1/2}\sqrt{\Gamma(\Gamma-1)} \cong 0.4 (\e_B n_0)^{1/2} \Gamma
 \cong 1200 \sqrt{\e_B n_2}\Gamma_{300}\;{\rm Gauss}
\end{equation}
Thus 
\begin{equation}
E_{max} \cong 8\times 10^{20} Z n_2^{1/6} \e_B^{1/2} \Gamma_{300}^{1/3} E_{54}^{1/3}\;{\rm eV}
\end{equation}
\cite{vie98,dh01}, 
so that external shocks of GRBs can accelerate particles 
to ultra-high and, indeed, super-GZK energies.
Implicit in this result is that acceleration occurs within the GRB blast wave through, 
for example, second-order Fermi acceleration \cite{dh01}. Acceleration to ultra-high energy 
through first-order relativistic shock acceleration requires a highly magnetized surrounding
medium \cite{ga99}.

\section{Rapid X-ray declines from UHECR escape}

If UHECRs are accelerated by GRB blast waves, then blast-wave dynamics
will be affected by the loss of internal energy when the UHECRs escape. 
This effect is proposed to explain the rapid X-ray declines in the Swift GRB light
curves \cite{der06b}. Photohadronic processes become important
when the threshold condition $\ep\gamma^\prime \gtrsim m_\pi/m_ec^2 \simeq
400$, where $\e = h\nu/m_ec^2$ is the dimensionless photon energy,
$m_pc^2 \gamma$ is the proton energy, and $\g$ is the proton Lorentz 
factor. For protons interacting with 
photons at the peak photon energy $\e_{pk}\cong 2\Gamma \ep_{pk}/(1+z)$ 
of the $\nu F_\nu$ spectrum, 
\begin{equation}
E_{pk} \simeq {3\times 10^{16} (\Gamma/300)^2\over (1+z) \e_{pk}}\;{\rm eV}.
\label{Epk}
\end{equation} 

The comoving timescale for a proton to lose a significant fraction of its energy 
through photohadronic processes is given by $t^\prime_{\phi\pi}(\gamma)$, 
where $t^{\prime -1}_{\phi\pi}(\gamma) \simeq (K_{\phi\pi}\sigma_{\phi\pi}) 
[\ep n^\prime_{ph}(\ep )] c$, $K_{\phi\pi}\sigma_{\phi\pi}$ $ \cong 70~\mu$b
is the product of the photohadronic cross section and 
inelasticity, and the comoving
energy density of photons with energy $\approx \ep$ is
$u_{\ep}^\prime \cong m_ec^2 \e^{\prime 2}n^\prime_{ph}(\ep )$.

The relation between the measured $\nu F\nu$ flux $f_\e$ and internal energy density 
is $u_{\ep}^\prime \cong d_L^2 f_\e/(c x^2 \Gamma^2)$, where $d_L = 10^{28} d_{28}$ cm is the 
luminosity distance of the GRB. For protons interacting
with photons with energy $\ep_{pk}$, we therefore find that the comoving time required
for a proton with energy $E_{pk}$ (as measured by an observer outside the blast wave) to lose
a significant fraction of its energy through photohadronic processes is
\begin{equation}
t^\prime_{\phi\pi}(E_{pk}) 
\simeq 
{m_ec^2 x^2 \Gamma^2 \ep_{pk}\over K_{\phi\pi}\sigma_{\phi\pi}d_L^2f_{\e_{pk}}}
\simeq 
2\times 10^6\; {x_{16}^2 (\Gamma/300) (1+z) \e_{pk}\over d_{28}^2 f_{-6}}\;{\rm s}\;,
\label{tprimephipi}
\end{equation} 
where $x = 10^{16} x_{16}$ cm and
 $f_{\e_{pk}} = 10^{-6}f_{-6}$ ergs cm$^{-2}$ s$^{-1}$ is the $\nu F_\nu$ flux measured at $\e_{pk}$; the relation between $E_{pk}$
and $\e_{pk}$ is given by eq.\ (\ref{Epk}).

The dependence of the 
terms $x(t)$, $f_{\e_{pk}}(t)$, $\Gamma(t)$, and $\e_{pk}(t)$ on 
observer time in eq.\ (\ref{tprimephipi}) 
can be analytically expressed for the external shock model 
in terms of the GRB blast wave properties $E_0$, $\Gamma_0$, 
environmental parameters, e.g., $n_0$, and 
microphysical blast wave parameters $\e_B$ and $\e_e$ \cite{der06b}. 
This can also be done for other important timescales, 
for example, the (available) comoving time $\tp_{ava}$
since the start of the GRB explosion, the comoving acceleration time 
$\tp_{acc} = \zeta_{acc} m_pc^2 \gp/eBc $, written as a factor $\zeta_{acc}\gg 1$ times 
the Larmor timescale \cite{rm98}, the escape timescale $\tp_{esc}$ in the Bohm
diffusion approximation, and the proton synchrotron energy loss timescale $\tp_{syn}$.

\begin{figure}
\includegraphics[width=32pc]{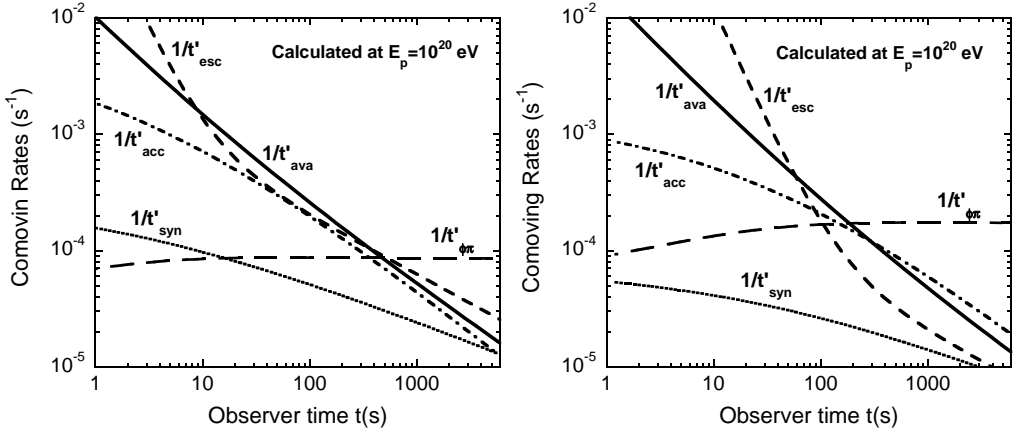}     % includes figure foo.eps
\caption{Rates and inverse timescales as a function of 
observer time for $10^{20}$ eV cosmic ray protons as measured 
by a stationary external observer. Left and right panels are
 results for parameter sets 1 and 2, respectively.}
\end{figure}

Fig.\ 1 shows the rates (or the inverse of the timescales) for 
$10^{20}$ eV protons in the case of 
an adiabatic blast wave that decelerates in a uniform
surrounding medium. The left-hand 
panel of Fig.\ 1 uses the parameter set
$$z = 1, \;\Gamma_{0} = 300, \;E_{54} = 1, \;n_0 = 1000{\rm~cm}^{-3}, \;\e_{e} = 0.3, \;\e_{B} = 0.3\;,$$ 
and the right-hand panel uses the parameter set
$$z = 1, \;\Gamma_{0} = 150,\; E_{54} = 10,\; n_0 = 1000{\rm~cm}^{-3}, \;\e_{e} = 0.1, \;\e_{B} = 0.3\;.$$
The characteristic deceleration timescale in the left and right
cases, given by  
$t_d \cong 
9.6(1+z) ( {E_{54}/ n_2 \Gamma_{300}^8})^{1/3}$ s,
is $\approx 9$ s and $\approx 120$ s, respectively.

For these parameters, it takes a few hundred seconds to accelerate
protons to energies $\approx 10^{20}$ eV, at which time photohadronic
losses and escape start to be important. Photohadronic losses inject 
electrons and photons into the GRB blast wave. The electromagnetic
cascade emission, in addition to hyperrelativistic electron 
synchrotron radiation from neutron escape followed by subsequent
photohadronic interactions \cite{da04}, makes a delayed anomalous 
$\gamma$-ray emission component
as observed in some GRBs \cite{hur94,gon03}. Ultra-high energy 
neutrino secondaries are produced by the photohadronic processes.
Detection of high-energy neutrinos from GRBs would confirm the importance of 
hadronic processes in GRB blast waves.
The ultra-high energy neutrons and escaping protons
form the UHECRs with energies $\gtrsim 10^{20}$ eV.  

The GRB blast wave rapidly loses internal energy due to the 
photohadronic processes and particle escape. The blast wave
will then rapidly decelerate, producing a rapidly decaying
X-ray flux. As argued in more detail elsewhere \cite{der06b},
the rapidly decaying fluxes in Swift GRBs are signatures of UHECR
acceleration by GRBs. 
If this scenario is correct, GLAST will detect anomalous 
$\gamma$-ray components, particularly in those GRBs that undergo rapid 
X-ray declines in their X-ray light curves.

\acknowledgments
This work is supported by the Office of Naval Research,
by NASA {\it GLAST} Science Investigation No.\ DPR-S-1563-Y, and 
NASA Swift Guest Investigator Grant No.\ DPR-NNG05ED41I.
Thanks also to Guido Chincarini for the kind invitation.

\end{document}